\documentclass[graybox,table]{svmult}

\usepackage[table,xcdraw]{xcolor}
\usepackage{doi}

\usepackage{type1cm}        
\usepackage{makeidx}         
\usepackage{graphicx}        
\usepackage{multicol}        
\usepackage[bottom]{footmisc}

\usepackage{newtxtext}       %
\usepackage[varvw]{newtxmath}       

\makeindex             

\usepackage[utf8]{inputenc}
\usepackage{geometry}
\usepackage{hyperref}
\usepackage{scrextend} 
\usepackage[table,xcdraw]{xcolor} 

\usepackage{natbib}

\usepackage{graphicx}
\usepackage[parfill]{parskip}
\usepackage{amsmath}
\usepackage{pifont}
\usepackage{wasysym}
\newcommand{\cmark}{\ding{51}}%
\usepackage{booktabs}
\usepackage{multirow}
\usepackage{adjustbox}
\usepackage{subcaption}
\usepackage{url}

\begin{document}

\title*{A Guide to Evaluating the Experience of Media and Arts Technology}

\author{Nick Bryan-Kinns and Courtney N. Reed}
\institute{Nick Bryan-Kinns \at Creative Computing Institute, University of the Arts London, London, UK, and \\Queen Mary University of London, Mile End, London, UK \email{n.bryankinns@arts.ac.uk}
\and Courtney N. Reed \at Queen Mary University of London, Mile End, London, UK, and \\Max Planck Institute for Informatics, Saarland Informatics Campus, Saarbrücken, Germany \email{creed@mpi-inf.mpg.de}}
%
%
\maketitle\vspace{-3ex}

\abstract{
Evaluation is essential to understanding the value that digital creativity brings to people’s experience, for example in terms of their enjoyment, creativity, and engagement.
There is a substantial body of research on how to design and evaluate interactive arts and digital creativity applications \cite{Candy14}. There is also extensive Human-Computer Interaction (HCI) literature on how to evaluate user interfaces and user experiences \cite{Blythe04}. However, it can be difficult for artists, practitioners, and researchers to navigate such a broad and disparate collection of materials when considering how to evaluate technology they create that is at the intersection of art and interaction. \\
This chapter provides a guide to designing robust user studies of creative applications at the intersection of art, technology and interaction, which we refer to as \emph{Media and Arts Technology} (MAT). We break MAT studies down into two main kinds: \emph{proof-of-concept} and \emph{comparative studies}. As MAT studies are exploratory in nature, their evaluation requires the collection and analysis of both qualitative data such as free text questionnaire responses, interviews, and observations, and also quantitative data such as questionnaires, number of interactions, and length of time spent interacting. This chapter draws on over 15 years of experience of designing and evaluating novel interactive systems to provide a concrete template on how to structure a study to evaluate MATs that is both rigorous and repeatable, and how to report study results that are publishable and accessible to a wide readership in art and science communities alike. 
}

\noindent\rule{4cm}{0.4pt}

\textbf{Preprint}. Chapter to appear in \emph{Creating Digitally. Shifting Boundaries: Arts and Technologies — Contemporary Applications and Concepts}, Anthony L. Brooks (Editor), Springer.\\ \url{https://link.springer.com/book/9783031313592}

\newpage

\section{Introduction}
\label{chap:intro}

Media and Arts Technology (MAT) research exists in an interdisciplinary space at the intersection of artistic and creative practice, technology innovation, and research on human cognition and interaction. {With MATs, the design of the creative application itself and the study of its use through user studies are intertwined and may take the form of a scientific intervention or an examination}. This poses a challenge to you, as a researcher: on one hand, it offers opportunities for novel and engaging exploratory research and yet, at the same time, needs to be undertaken and framed in a way that works within current academic research discourse and vernacular. {There is a natural tension between the focus of science and the arts when examined independently: {science tends to focus on building and testing generalisable models and adding knowledge to our understanding of the world} \citep{England2016,Edmonds2011InteractingAR}. 
Artistic practice approaches the understanding of the world slightly differently, often focusing toward creativity and the individuality and subjectivity of the human condition, whether or not that produces any particularly novel understanding of the world \citep{Edmonds2011InteractingAR, Candy2018}. However, these fields are intertwined and inseparable, having encouraged each other's advancement since the earliest ventures in philosophy and understanding of the world \citep{FelipeDuarte2019, Benford2013}.}
In order to conduct quality research in MAT, we must acknowledge both components: you as the researcher are responsible for honoring the inventiveness, innovation, and adaptability of the arts in a way which adheres to the structure and procedure of scientific research. 

\subsection{Principles of Quality Research}
In order to make your work understandable to, and valued across research communities, it should meet two critical requirements: it must be both \textbf{rigorous} and \textbf{repeatable}:

\textbf{Rigorous} means that your research is conducted using tried and tested scientific methods and practices to collect and analyse data. MAT research is interdisciplinary, exploring new forms of digital \underline{M}edia and \underline{A}rts, and yet is rooted in the science and engineering of the \underline{T}echnology. With these fields being extremely broad in their own right, MAT research must be diligent in conducting research that is appropriate for, and consistent with, existing bodies of knowledge. It is therefore vital to collect and analyse data in a structured way using established methodologies from scientific fields, to ensure that data collected and results analysed are considered to be valid and reliable by other researchers.

\textbf{Repeatable} means that your study plan and methodology are written in enough detail that someone else could run the study without you being present to explain it. This is important to maintain consistency of your study (each time you run the study it is done the same way), and to allow others to be able to reproduce your results if they want to.

Keeping these two core requirements in mind, you can design and evaluate compelling research which use MATs to contribute to the different fields in interdisciplinary research. The goal of this chapter is to outline practices for conducting and presenting research in a rigorous and repeatable way. In addition to providing guidelines for structuring your research, we illustrate these through examples of sound research practices used in existing MAT research.

This chapter starts by introducing the two types of MAT study --- proof-of-concept and comparative --- and provides example studies of both varieties, {ranging from interactive music technology to playful tangible interaction}. Referencing these existing projects, {we introduce a series of guidelines for designing MATs as part of an user study and explain how to define relevant research questions for your research. Then, we outline methods to determine appropriate data to collect from quantitative and qualitative sources and how these are combined as `mixed methods'. A guide to questionnaire and interview design is provided, which focuses on how to elicit further insights on experience from participants. Tools and techniques for qualitative and quantitative data analysis are introduced, including thematic analysis \citep{braun:2012} for interview data and statistical analysis for questionnaire and interaction data. 
The chapter concludes with guidance on how to report the results of the user studies so that others may be able to refer to your work and findings for their own design and technology.}

{In implementing the robust study design practices discussed in this paper, we hope that MAT researchers will find more common ground and understanding of each other's multidisciplinary work. As such, this chapter serves as a guide for researchers from many disciplines studying MATs, leading to richer collaboration and dissemination of knowledge across research communities.}

\subsection{HCI, UX, and Interactive Arts}
Before moving on, we would like to briefly discuss the historical link between the more computer science fields of HCI and user experience (UX) and digital arts, media, and creativity. The development of these fields has always been tightly intertwined \citep{England2016, Nam2013}, with the arts providing a source of inspiration for technology and creativity in its evaluation \citep{FelipeDuarte2019}. 
The earliest uses of computers to create art were in the 1950s, for example, in 1951 one of the earliest example of computer arts was the use of the Ferranti Mark 1 computer at the University of Manchester (there would only have been a handful of computers in the UK at the time) to play simple tunes. 
At a similar time Human Factors and Ergonomics, which are the origins of contemporary HCI and UX, emerged in the 1940s and 1950s to help designers design machines which were easier to use, for example, to improve the design of airplane controls to make flying safer and less error prone. The focus for Human Factors was really on the functionality of the machine. A lot of it was to do with how to layout, or configure, the controls in order to reduce chance of human error.
This relationship is reciprocal and has worked in a collaborative way over the decades --- typically, technology from other fields is adopted into arts, where it is used creatively. From these applications, new practices are developed and learned, stimulating the study and further expanding the technology itself. Interactive Art tends to place importance on what the user or listener cares about \citep{FelipeDuarte2019, Jeon2019}, as well as challenging the status quo to provide room for novel ideas \citep{Benford2013}. This provides a space for HCI to move beyond more traditional HCI topics such as task completion and efficiency to understanding and communication between humans and computational agents \citep{Jeon2019}.

\section{Types of MAT Study}
\label{chap:ex}

There are two broad types of MAT study which we will discuss in this chapter:

\textbf{Proof-of-concept} studies examine people’s responses to a single MAT. This kind of study asks \textit{“What if...”} questions such as “What if I make this MAT, how do people respond to it, and what is their experience of it?” In this way, these studies often involve creating a MAT which is a digital intervention and then evaluating people's response to it. These studies allow early-stage investigation of how people respond to a new form of interaction, for instance when there is little existing research in the area and it is therefore unclear how people might respond to the interaction. Results of such a study can inform further studies by identifying broad kinds of response to the MAT and identifying possible interesting and challenging avenues for interaction design.

\textbf{Comparative studies} do as their name indicates -- they compare two or more variations of a particular MAT. 
These studies compare the effects of specific design features and ask \textit{“What effect do particular design feature(s) have on people’s experience?”} Rather than trying to determine whether one design is better than another, these studies are interested in \emph{how} the experience is different between MATs. In this way, comparative studies involve in-depth examination of MATs and differences are typically restricted to one design feature. Results inform the development of theories about how design features might contribute to enjoyment, creativity, and engagement, and generate interaction design questions for future studies. 

Often, proof-of-concept studies are used to research a general interaction question, which is then further refined and explored through specific questions in a comparative study. However, this is not necessarily the case and it should be made clear that one kind of study is not ``better'' or more rigorous or repeatable than the other -- they simply ask different kinds of questions. As Section \ref{chap:design} will elaborate, it is important to keep your research goals in mind in order to decide and justify the methods you use.  

In this chapter, papers published about existing MAT research projects are used to illustrate and bring to life these different study types, how they can be conducted, and how they are presented. These projects are briefly introduced in the following subsections to give a flavour for different kinds of MATs and user study approaches.

\subsection{Example Proof-of-Concept Studies}
The proof-of-concept studies we refer to throughout this chapter are very different from one another but share the same type of research focus and question: a MAT is designed in order to conduct an open-ended exploration of users' behaviour and interaction. As mentioned above, although the research questions are very specific, they ask more of \textit{``What if?''} or \textit{``How do...?''} questions and are aimed at exploring the general effects of a MAT's use as illustrated below.

\subsubsection{Mazi}
\citet{nonnis:2019} designed and studied Mazi, a Tangible User Interface (TUI) which uses haptic and auditory feedback to encourage spontaneous and collaborative play between children with high support needs. Mazi was developed through an iterative prototyping process and used in a proof-of-concept style study to explore how principles of TUI design along with theories of social interaction could be used to encourage social play \citep{Nonnis:2019:IDC}. Mazi's final design features a dome-like shape to facilitate the circular configurations found naturally in communicative behavior and uses soft yet durable materials to allow the children to play in their own way with sensors embedded in Mazi which generate music. The core proof-of-concept research question of Mazi was: 

\begin{addmargin}[1em]{2em}
\emph{
``\textbf{What if} I make this MAT a large, soft circular shape and make it create music when multiple children play with it at the same time, \textbf{how do} children with autism respond to this, and what playful and social interaction does it prompt?''.  
}

\end{addmargin}

Because the authors wanted to determine how the tangible and auditory feedback of the design influenced the communication between the children, it was most suitable to conduct a proof-of-concept study; indeed, the wide variety of abilities and interests of children involved makes it unsuitable for comparative study. The proof-of-concept study took place over five weeks with five children aged between 6 and 9 years old at a Special Education Needs (SEN) in London.

The study process needed to be flexible to the needs of the children and teachers. In the proof-of-concept study, the researchers collected observational data about how the children interacted with each other and with Mazi, and analysed these using existing behavioural science models. The results of the study demonstrated that working with Mazi helped the children to master basic social skills and engage with different sensory interactions \citep{Nonnis:2019:IDC}. This open-ended strategy allowed the researchers to focus on the most salient elements of the interaction in-context, and yet at the same time produced a study method and results which could be replicated by other researchers. 

\begin{figure}[h!]
     \centering
     \begin{subfigure}[b]{0.49\textwidth}
         \centering
            \includegraphics[width=\textwidth]{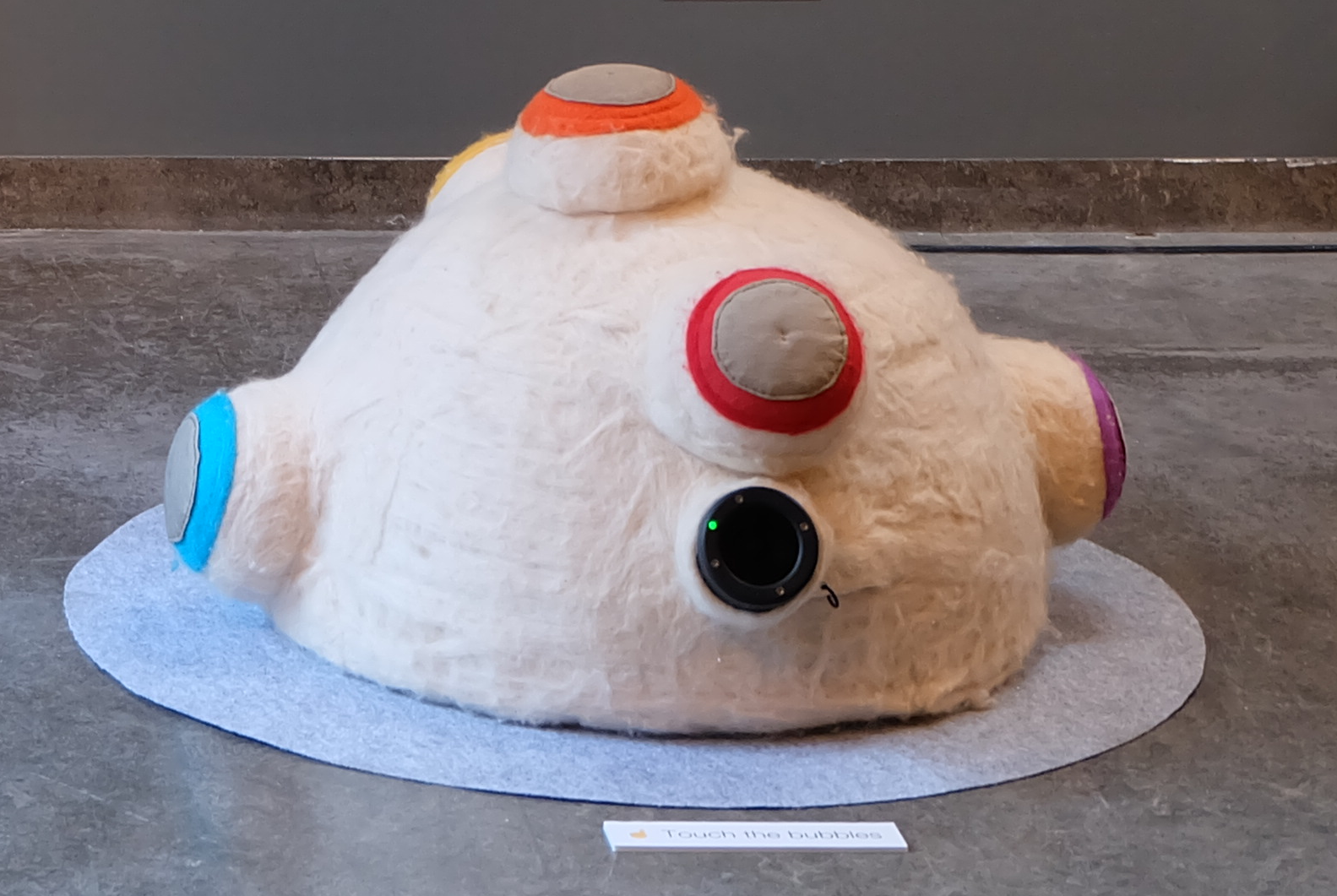}
            \caption{Mazi, made of wool and featuring inflatable bubbles for triggering sounds.}
         \label{fig:temp-a}
     \end{subfigure}
     \hfill
     \begin{subfigure}[b]{0.47\textwidth}
         \centering
            \includegraphics[width=\linewidth]{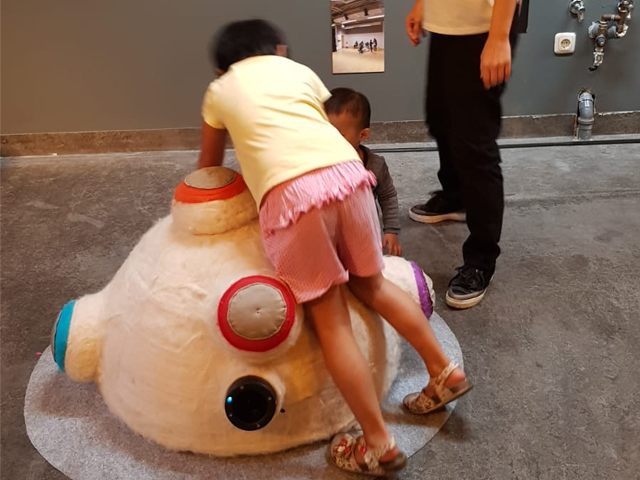}
            \caption{Children playing together with Mazi.}
         \label{fig:temp-b}
     \end{subfigure}
        \caption{Mazi, a tangible user interface for stimulating interaction and participation for autistic children. Images used with permission from the Authors: \href{http://isam.eecs.qmul.ac.uk/projects/Mazi/mazi.html}{http://isam.eecs.qmul.ac.uk/projects/Mazi/mazi.html}}
        \label{fig:temp}
\end{figure}

\subsubsection{VoxEMG}
\begin{figure}[h!]
     \centering
     \begin{subfigure}[b]{0.47\textwidth}
         \centering
            \includegraphics[width=\textwidth]{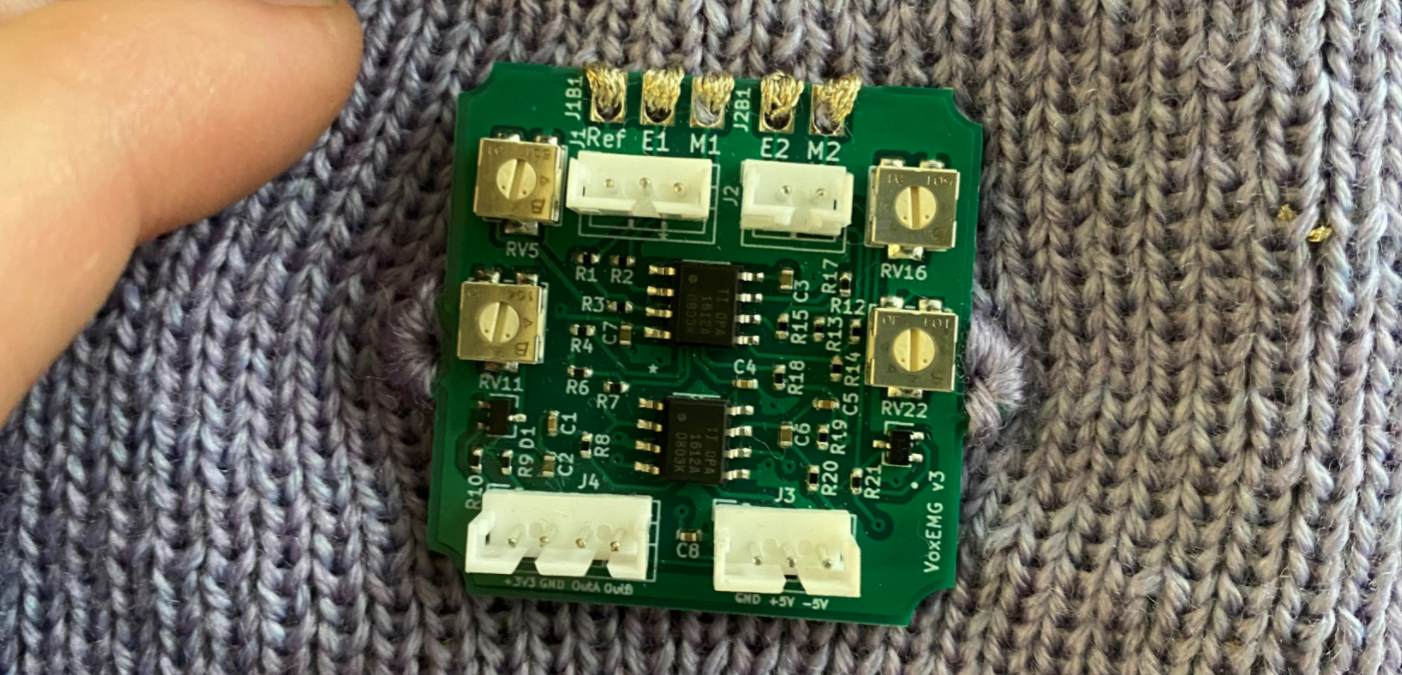}
            \caption{The VoxEMG board developed for sensing activation of the laryngeal muscles.}
         \label{fig:voxEMG-1}
     \end{subfigure}
     \hfill
     \begin{subfigure}[b]{0.49\textwidth}
         \centering
            \includegraphics[width=\linewidth]{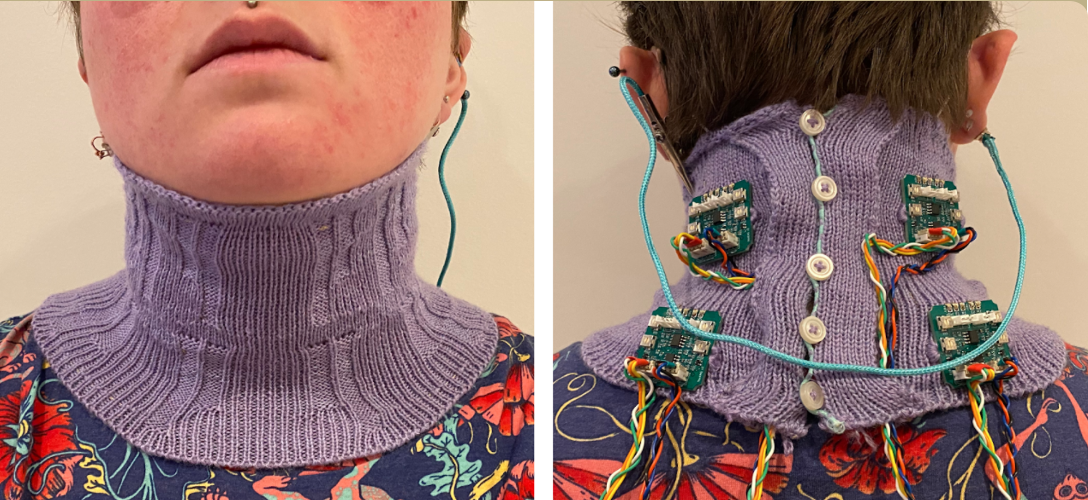}
            \caption{Reed wearing the Singing Knit wearable collar for vocal EMG interaction.}
         \label{fig:voxEMG-2}
     \end{subfigure}
        \caption{Sensing and interacting with laryngeal muscular activations through VoxEMG integrated into wearable designs (reproduced from \citet{reed:singing-knit}).}
        \label{fig:voxEMG}
\end{figure}

\citet{reed:direct-vocal-control} design and explore the use of novel vocal interaction through surface electromyography (EMG) using an autobiographical approach \citep{Neustaedter:2012:autobiographical}. The authors developed a system, the VoxEMG, for gathering the electrical neural impulses which cause muscular contractions (Figure \ref{fig:voxEMG}). These EMG signals are used in real-time sound design to allow a singer to interact with very low-level movements in their practice through auditory feedback. The auditory feedback allows the vocalists to interact with their existing, embodied understanding of their action and ``hear'' movements which would not normally produce definable sound. The work was formed through an autobiographical approach in Reed's extended experience with the setup lasting over a year. The fundamental research question was:

\begin{addmargin}[1em]{2em}
\emph{
``\textbf{What if} I sonify movements and actions which singers are not normally consciously aware of, and \textbf{how will} they react, change, and/or perceive their movements when they receive this new feedback?''.
}
\end{addmargin}

Autobiographical methods and first-person accounts are extremely useful as they demonstrate how lived-experience and understanding of a system's use can improve and inform its design \citep{hook:2015:1PPsomadesign}. In this case, Reed applied her experience working as a semi-professional vocalist to the long-term interaction with the EMG system. Through detailed interaction notes, journaling, debugging, and an iterative design and testing process, Reed was able to uncover subtle understanding of the interactions with it \citep{Neustaedter:2012:autobiographical, reed:2021}. This study again is more suitable as a proof-of-concept because it requires an exploratory approach; although there are some hypotheses as to how the musicians would respond, the researchers wanted to keep the interaction open-ended and see how the singer's behaviour changed while using this MAT.

\subsubsection{Polymetros}
\begin{figure}[ht]
     \centering
     \begin{subfigure}[b]{0.68\textwidth}
         \centering
            \includegraphics[width=\textwidth]{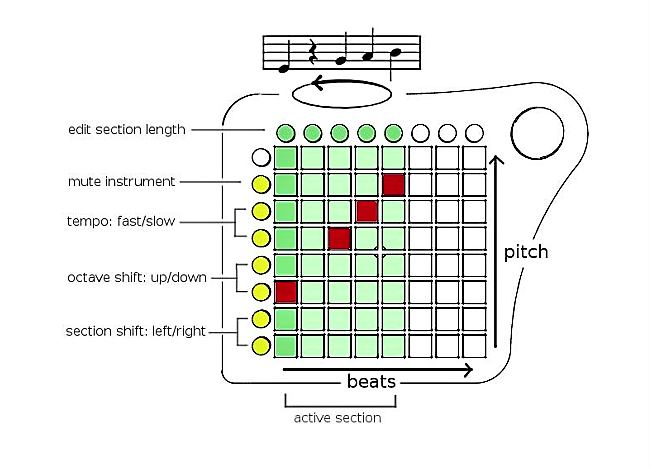}
            \caption{The layout of each Polymetros instrument.}
         \label{fig:polymetros-1}
     \end{subfigure}
     \hfill
     \begin{subfigure}[b]{0.58\textwidth}
         \centering
            \includegraphics[width=\linewidth]{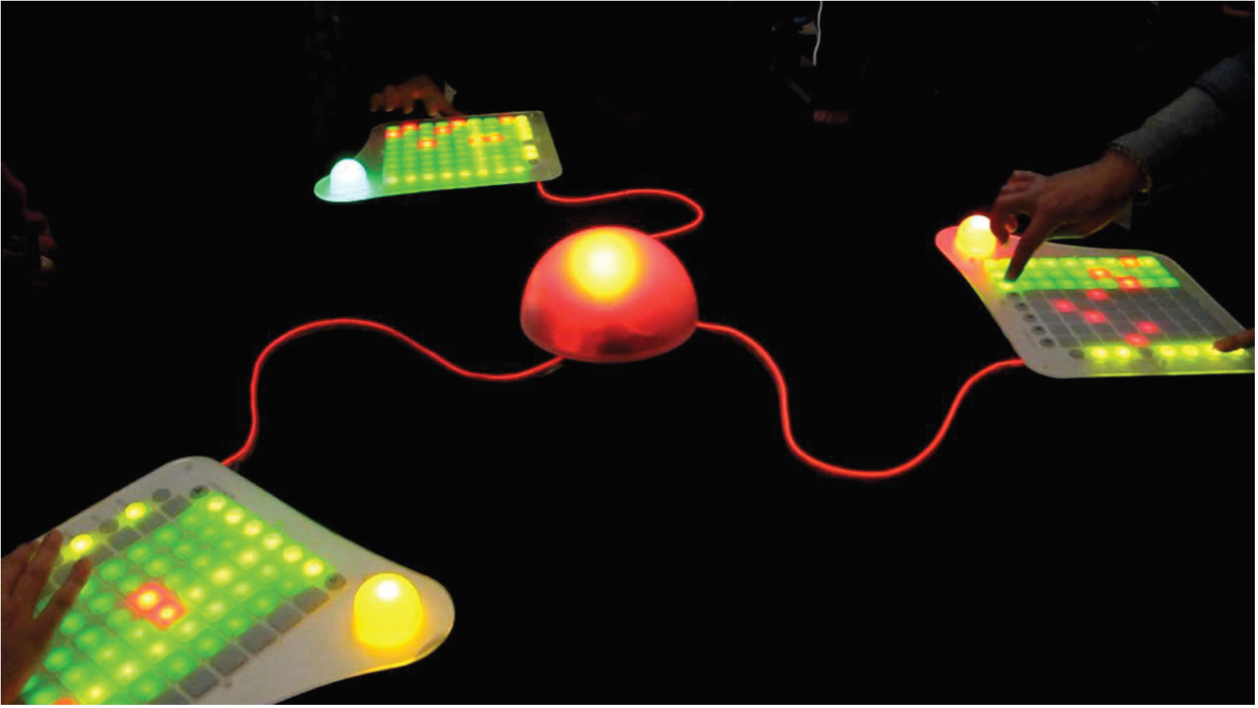}
            \caption{Creating collaboratively between three musicians with the Polymetros system.}
         \label{fig:polymetros-2}
     \end{subfigure}
        \caption{The Polymetros collaborative music system (reproduced from \citet{bengler:2013}).}
        \label{fig:polymetros}
\end{figure}

Polymetros is a collaborative music system designed as an in-person audience experience for multiple participants (Figure \ref{fig:polymetros}) \citep{bengler:2013}. The design is purposefully simple, using minimal music, and allowing participants to create short loops (8 notes long with 8 possible pitches and only one instrument sound). There are 3 physical instruments in Polymetros (Figure \ref{fig:polymetros-1}) which are synchronised together so that the loops of each instrument are synchronised with each other. One person can play Polymetros on their own, but the sound becomes richer and more interesting as 2 or 3 instruments are played at the same time (Figure \ref{fig:polymetros-2}). Essentially the research question of this MAT was:

\begin{addmargin}[1em]{2em}
\emph{
``\textbf{What if} I make a MAT musical instrument which requires three people to play it, \textbf{how do} people respond to it, and what is their experience of it?''
}
\end{addmargin}

\citeauthor{bengler:2013} wanted to explore how people would respond to this novel collaborative music system, how they would engage with it, and how they might engage with each other. With these open-ended questions, a proof-of-concept study is the most appropriate. A study was conducted over two days at the Victoria \& Albert Museum with random members of the public (150+), to see how they would respond to the novel form of music making. Data was collected with many different tools, including questionnaires, observations, video recordings, and data logs from the Polymetros system. Later studies then took on a more comparative study structure, comparing how Polymetros was perceived by people in different cultural contexts - in the UK, Spain, and China \citep{Bengler2014}. Results indicate the significance of ownership and supporting individual participation in collaborative creativity; the physicality of the system's design assisted in non-verbal communication and understanding of the other players' actions and structure roles in the compositional process \citep{bengler:2013}. 

\subsection{Comparative Studies}
In contrast, the comparative studies we will discuss in this chapter are focused on particular aspects of interaction or effects of the MAT's use. These studies are different in the questions they ask, which are more focused on ``How does X effect Y'' or ``Can we do...''. They focus on the design of an interactive system and specific design features can change or impact users' interaction as introduced in the following examples.

\subsubsection{Keppi}
\begin{figure}[ht]
         \centering
            \includegraphics[width=0.5\textwidth]{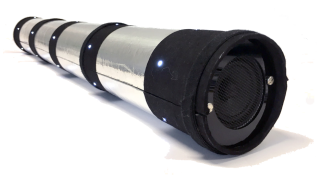}
        \caption{The Keppi, a cylindrical instrument designed with intentional disfluency characteristics (reproduced from \citet{bin:2018}, with permission).}
        \label{fig:keppi}
\end{figure}

Keppi \citep{bin:2018} is a Digital Musical Instrument (DMI) designed to explore the effect that disfluency in a musical instrument's design might have on performers' and audiences' perception of skill and risk in performance (Figure \ref{fig:keppi}). The research question here started as a more open question of ``What if I make this MAT musical instrument which is risky to play in performance as its musical properties degrade in real time'' and then moved to a more focused Comparative Study question: 

\begin{addmargin}[1em]{2em}
\emph{``\textbf{What effect} does increasing the disfluency in the design of a MAT musical instrument have on audience and performer perception of skill and risk in live performance?'' 
}
\end{addmargin}

In this comparative study Keppi was designed and produced incorporating a disfluent design characteristic: It would turn itself off if not constantly moved. Six percussionists then performed live on stage with different versions of Keppi which each had different levels of disfluency. Audience feedback was collected in real time during the performances through an app on their mobile phones, and also through post-event questionnaires. Performer feedback was collected through survey questions, and the style of music created and performed using Keppi was analysed. The results of the study suggested that whilst different levels of disfluency in the design of Keppi did not have an effect on audience enjoyment of the performance, it did have an effect on their recognition of the skill of the performers. Moreover, performers noted that the disfluent behaviour of Keppi was viewed as a positive design feature, which contradicts conventional Human-Computer Interaction design guidelines which stress the importance of intuitive and reliable user interfaces.

\subsubsection{Daisyphone}
\begin{figure}[ht]
     \centering
     \begin{subfigure}[b]{0.62\textwidth}
         \centering
            \includegraphics[width=\textwidth]{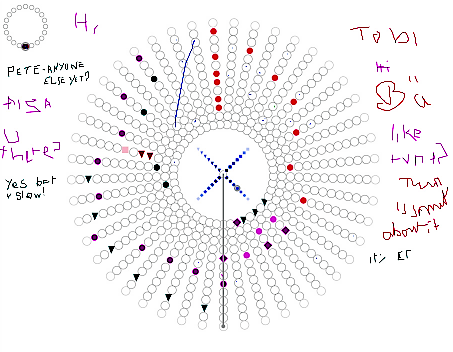}
            \caption{The Daisyphone interface on the desktop, showing player input and annotations.}
         \label{fig:daisyphone-1}
     \end{subfigure}
     \hfill
     \begin{subfigure}[b]{0.64\textwidth}
         \centering
            \includegraphics[width=\linewidth]{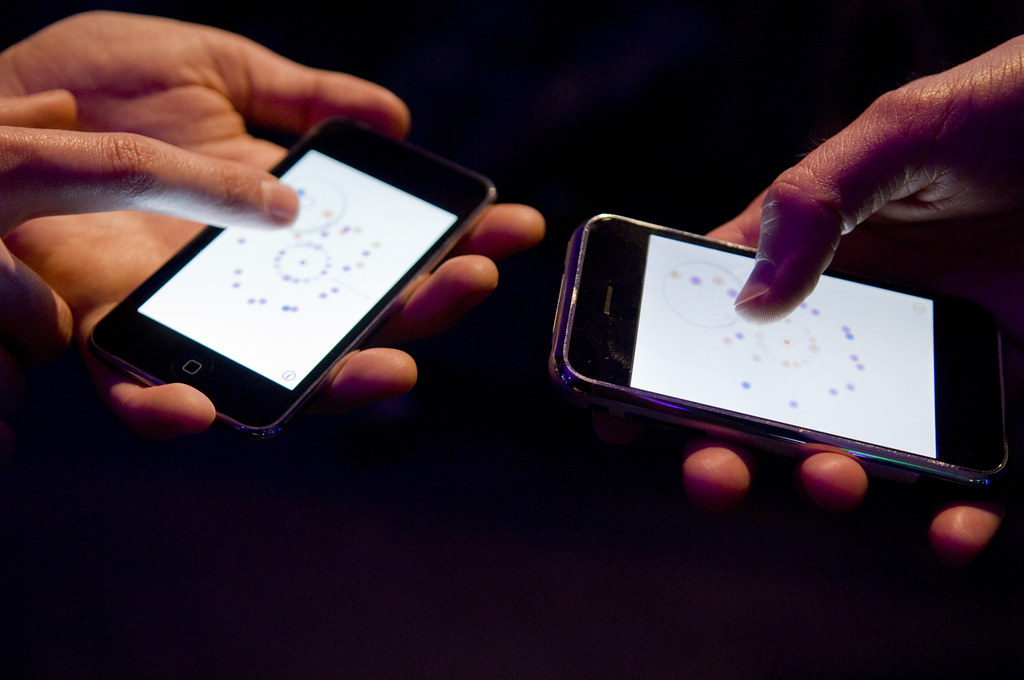}
            \caption{Compositional collaboration with Daisyphone on mobile.}
         \label{fig:daisyphone-2}
     \end{subfigure}
        \caption{The Daisyphone interface for collaborative music-making (Fig. \ref{fig:daisyphone-1} reproduced from \citet{bryankinns:2009}, Fig. \ref{fig:daisyphone-2} \copyright EPSRC, available to the public: \href{https://www.flickr.com/photos/epsrc/3340524049/}{https://www.flickr.com/photos/epsrc/3340524049/}).}
        \label{fig:daisyphone}
\end{figure}

Daisyphone is an online collaborative music editor which allows groups of people to edit a shared loop of music 48 notes long (Figure \ref{fig:daisyphone}). The design is purposefully very simple, and the interaction is restricted to adding notes (12 pitches are possible, and 4 instrument sounds) and removing notes, and drawing to communicate \citep{bryankinns:2009}. The circular area shows the shared musical loop and drawn annotations can be seen around the outside. The circular representation of the loop was chosen purposefully to be different to most music sequencers, thereby reducing familiarity with the interface. Support for shared drawing and shared editing was drawn from HCI research which stated that this would improve collaboration (in-text document editing). \citet{bryankinns:2009} explored the effects of having a sense of personal identity and a shared way to communicate would have on people’s mutual engagement \citep{bryankinns:daisyphone}.
Mutual engagement being “it involves engagement with both the products of an activity and with the others who are contributing to those products” (ibid.). The research then focused around the question of:

\begin{addmargin}[1em]{2em}
\emph{
\textbf{What role} does personal identity play in collaborative musical composition, and \textbf{how do} different representations of personal identity \textbf{compare} when users complete a specific task together?
}
\end{addmargin}

The authors undertook a Comparative Study to see what effect providing cues to personal identity in the interface and providing a shared area to communicate in would have on people’s mutual engagement. The study involved 39 participants collaborating online (participants could not see or hear each other in person) to create short loops of music. Each participant spent about 1 hour in the study. Data was collected from questionnaires, and logs of interaction with both the music and the shared drawing area. Questionnaires were used to gather feedback from participants and to make comparisons between different versions of the Daisyphone interface.

\subsubsection{Smart Trousers}

\begin{figure}[ht]
         \centering
            \includegraphics[width=0.7\textwidth]{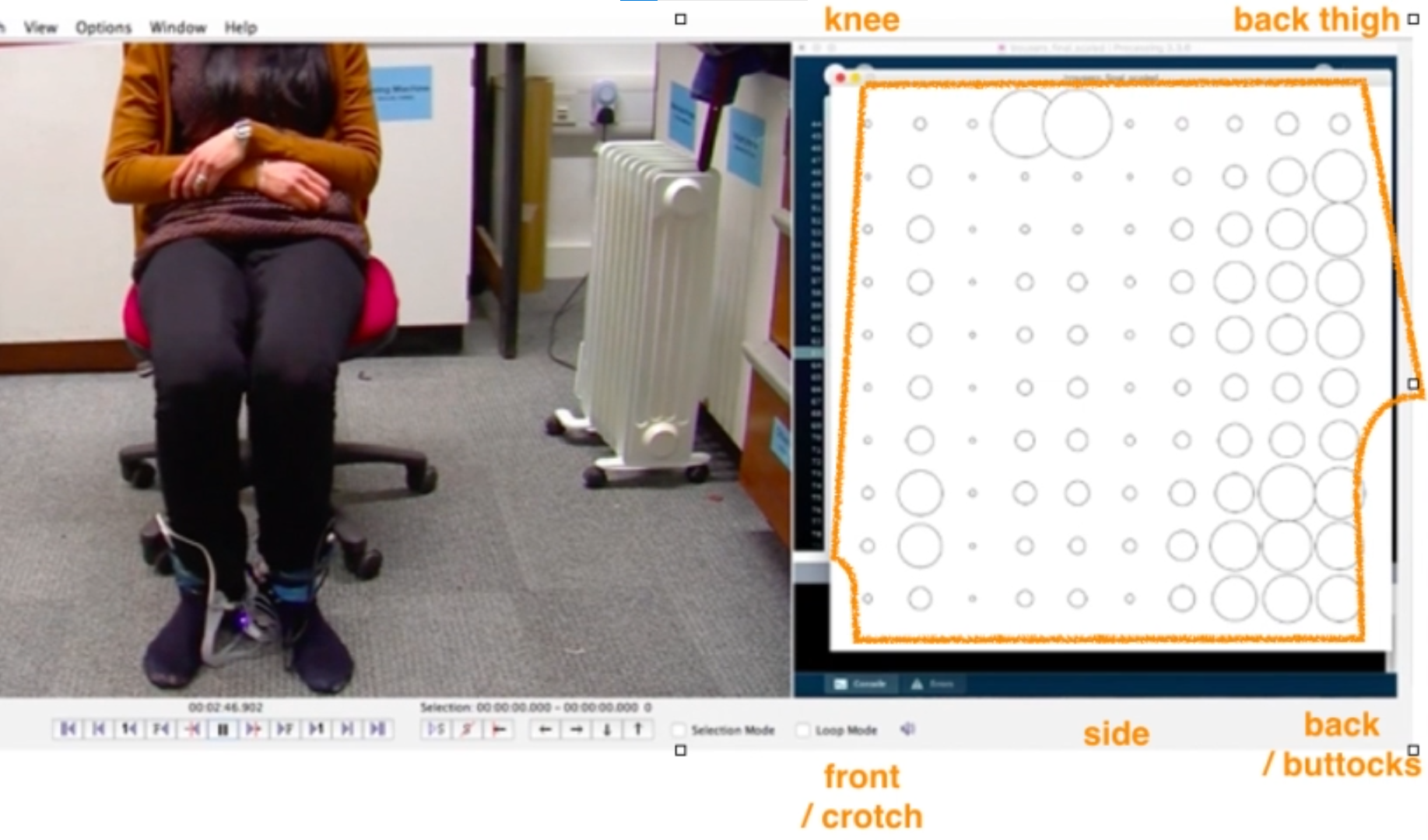}
        \caption{Visualising the pressure measurements (right) in a participant's seated position (left) while wearing the smart trousers (reproduced from \citet{skach:2018}), with permission.}
        \label{fig:trousers}
\end{figure}

\citet{skach:2018} designed a pair of smart trousers which were able to sense posture changes (Figure \ref{fig:trousers}). This MAT involved an interdisciplinary approach combining electronics and e-textile design with behavioural and social science. The use of a wearable and the authors' relevant fashion design background allowed for the creation of a wearable that could be used to study wearers' behaviour without being disruptive to a social environment. Fabric pressure sensors were integrated into the garment and measured contact between points on the wearer's body and the legs as well as the surface of a chair. Through this garment, the authors aimed to explore different behaviour related to emotional and social communication. The study focused on exploring non-verbal communication through posture and gestures and inquired specifically as to whether gestures on the legs could be classified through pressure sensing. The research was based around a question of:

\begin{addmargin}[1em]{2em}
\emph{
\textbf{Can we} use pressure sensing to gather data about the movement of the lower body when seated, and \textbf{can this data be used} to classify behaviours of the wearer as they engage in conversation?
}
\end{addmargin}

This work used a Comparative Study to gather data about participants wearing the smart trousers while they performed a number of actions. In this sense, the comparison is not between one MAT and another, but rather between different users and the same wearable. The interaction with the trousers showed that, even though individual participants followed the movement directions differently, in their individual interpretation, it was possible to classify different gestures using the data gathered. 

\subsection{From Proof-of-Concept to Comparative Study: The Chaos Bells}
Research into the interaction with large musical instruments by \citet{mice:large-instruments} provides an excellent example of how exploratory proof-of-concept studies can inform further comparative research.

This work began with a proof-of-concept interview study which explored musicians' relationship with the physically large instruments they had already been trained on \citep{mice:large-instruments}. Interview questions focused on gestures and the precision of movements, fatigue during performance, and improvisation. In addition, participants reviewed \textit{Cello Suite
no. 1 in G Major} (J. S. Bach) and discussed difficulties they would have playing the piece transposed for their own instrument. This exploration revealed a number of insights into timbral control and the embodied relationships between the musicians and their large instruments. 

\begin{figure}[ht]
     \centering
     \begin{subfigure}[b]{0.45\textwidth}
         \centering
            \includegraphics[width=\textwidth]{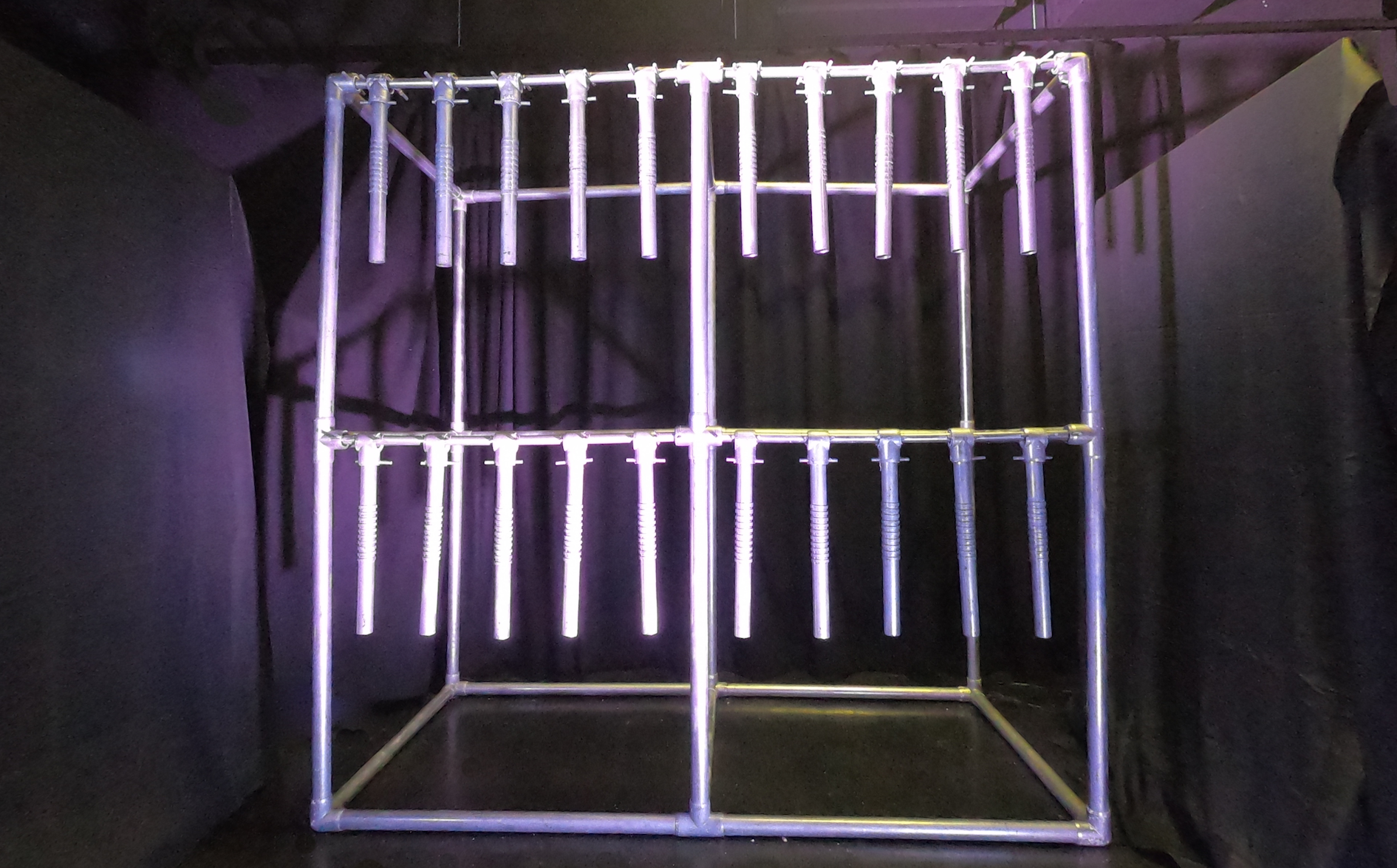}
            \caption{The Chaos Bells.}
         \label{fig:chaos-a}
     \end{subfigure}
     \hfill
     \begin{subfigure}[b]{0.51\textwidth}
         \centering
            \includegraphics[width=\linewidth]{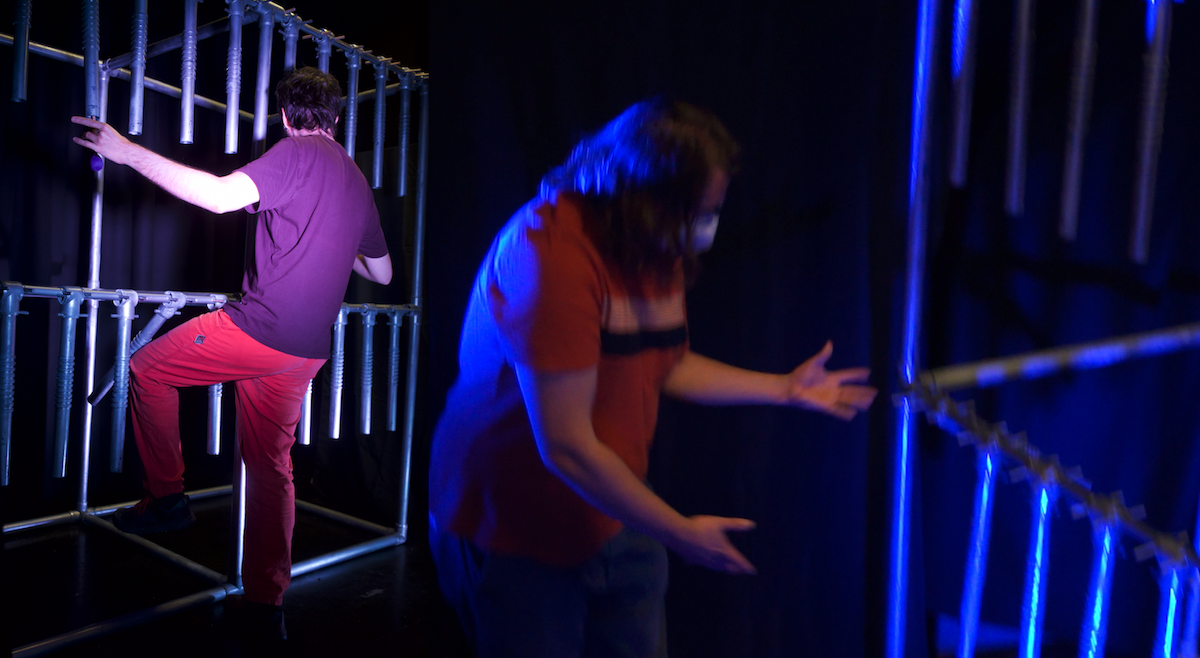}
            \caption{Performer gestures working with the large instrument.}
         \label{fig:chaos-b}
     \end{subfigure}
        \caption{The Chaos Bells, with height and width of 2 metres, designed to study the impact of interface size on interaction and gestures. Photos \copyright Lia Mice, used with permission.}
        \label{fig:chaosbells}
\end{figure}

From this study, the authors conducted a series of comparative studies using the Chaos Bells (Figure \ref{fig:chaosbells}), a large-scale digital musical instrument (DMI) developed by Mice \citep{mice:miming}. The design features a set of pendulums which use accelerometers which drive a Karplus-Strong algorithm as the performer strikes, raises, and swings them (Figure \ref{fig:chaos-a}). In further study of the instrument, the authors compare different pitch mappings on the pendulums to compare how different tonal layouts \citep{mice:m-in-nime, mice:miming}, instrument size, and the performer's body influence the idomatic gestures and patterns during improvisation, as well as the performers' perspectives of their own bodies \citep{mice:supersize} (Figure \ref{fig:chaos-b}), which show that the size of the instrument determines the gestures which can be used.

Through this work, we can see how the different study types can address different types of research questions using different approaches. The authors move between different focuses:

\begin{addmargin}[1em]{2em}
Q1: \emph{\textbf{How do} musicians perceive their interaction with their large instruments and \textbf{how might they act/ feel} about performing different music when playing them?} (proof-of-concept)

Q2: \emph{\textbf{Do different layouts} of the tones on a large instrument impact the gestures and movements used by musicians while playing them, and \textbf{does this change} their perception in practice?} (comparative)
\end{addmargin}

The proof-of-concept study used more open-ended approaches to gather a set of feedback and ideas surrounding the design and performance with large instruments. This was done to get a better sense of how performers work with instruments they have been trained on and know well. By using an exploratory approach, the authors determined key factors in the interaction with such instruments. This information informed the development of a new DMI based around these principles, where more specific questions of interaction such as tonal layouts could be examined in an appropriate context.

\section{Designing a MAT Study}
\label{chap:design}

\label{sec:research-qs}

The first step in conducting a robust MAT study is defining your research questions -- Having specific questions in mind will help to structure the rest of the research into either a proof-of-concept or comparative study. As discussed in Section \ref{chap:ex}, both kinds of study are equally valid but will focus on different kinds of questions and scenarios.
Having clear questions will drive consistency in the study and its analysis and focus the reporting of your results in a way which is understandable to others. This will also help to justify your design choices for both the study procedure, analysis, and the MAT --- {attention must be given to details of the \underline{M}edia and \underline{A}rts themselves (e.g., musical theory, visual arts practices, traditional craft) as well as the \underline{T}echnology driving the MAT interaction. The media and art should be researched in detail before beginning. For instance, when working with wearable MATs, it would be important to familiarise yourself with artists' practices in fashion design, wearable technology and integration, and so on. In the presentation of both Skach et al.'s smart trousers \cite[p.~117]{skach:2018} and Reed et al.'s VoxEMG wearable implementation \cite[pp.~171-172]{reed:singing-knit}, significant consideration and direction was first drawn from previous work in textile sensing technology. The design and use of the MAT will very likely rely on background and existing work from other science backgrounds as well, in these cases fields such as cognitive and behavioural sciences, bio-mechanics, and sensorimotor interaction.}

\subsection{Choosing a Study Type}

If your research interests are more exploratory, for instance wanting to know how people might interact with a new MAT in their typical performance practices (as done with VoxEMG and the Chaos Bells), proof-of-concept studies can provide a suitable format for your research questions. Similarly, proof-of-concept studies are good for questions such as \textit{How do people collaborate with this MAT?} \textit{How do participants' perceptions of their movement change if they use this MAT over a long time period?} \textit{How do participants choose which notes to play/ which sounds to include/ which roles to take on in a duet performance?} These questions are specific, but they do not seek to understand a specific difference -- rather, they are open-ended and aim to gather information about a specific context. This can be beneficial as well if you do not have any existing knowledge or want to prompt any particular behaviours. Proof-of-concept studies can be useful to observe the impact of and attitude towards a new design in-context -- for instance  Mazi worked with the children and their routines in their day-to-day environment \citep{nonnis:2019}.

On the other hand, research questions that directly involve comparisons naturally warrant a comparative study. For instance, the Daisyphone study's questions revolved around a comparison of features included in an user interface. Comparative studies are more appropriate if your questions are something like \textit{Do participants prefer one collaboration method over another?} \textit{Which of these interface features is more important to experienced users, compared to novice users?} or \textit{How does participants' accuracy score change after using this MAT for a month?} These research questions compare specific elements of experience and design and maybe compare responses from two or more groups of people (e.g., experts and novices). Additionally, you may be looking to gather information about different, specific moments in an interaction; for instance, with the smart trousers, examining the data gathered by the sensors at different moments in the interaction.

\subsection{Designing the MAT Itself}\label{sec:design-tech}
With these research questions in mind, you can decide on what MAT you will create and then how you will study it. When designing a MAT, it is also important to consider again that the methods are rigorous and repeatable -- someone else should be able to understand and recreate your design. With the growing movement of open-science, many researchers working in fields such as HCI are making their designs open-source so that they can be easily shared and accessed by other researchers and communities that will benefit from their use. Regardless of whether or not the MAT design is to be open-source, it is important to carefully document the design process to support the research in publication. Some key elements which should be included in this design documentation include:

\textbf{Design Inspiration}: MATs are results of inspiration-led design, often combining ideas and theories from different disciplines, and often the design focus is driven by from personal interest. For example, the collaborative music making system Daisyphone \citep{bryankinns:2009} was created as a result of a personal passion for performing music in combination with an interest in exploring new forms of networked music performance which were emerging at the time. If you are studying the design process or working in an autobiographical use case, you should note important details about your own experience and background which are relevant to the design (e.g., with VoxEMG \citep{reed:2021}).

\textbf{Identify Design Features}: Key design features of your MAT should be informed by previous research -- your study is likely to explore a new and novel context, but it should be based in established knowledge from the research community. For example, the design of Daisyphone was informed by Computer Supported Collaborative Work (CSCW) research, which stated that sense of identity was important in collaboration; therefore, identity was represented by colour in the user interface. This one design feature was then varied across different versions of the MAT to understand how sense of identity affected people’s collaborative creativity and engagement. The research questions therefore focused on this particular aspect.

\textbf{Reduce Complexity}: The complexity of the interaction design should be reduced as much as possible whilst still allowing for fun, creative, and engaging interaction. This makes the MAT easier to design and build and also makes the study more focused on the novel interaction itself rather than, say, learning effects. For example, Mazi was designed to support only a very small range of musical notes making it more of a ‘sound toy’ than a collaborative composition system. This allowed the studies to focus on how children interacted with each other rather than focusing on, say, the composition process. Additionally, keeping the design focused on novel interaction, rather than other elements of the design, will also help to reduce confounding factors ---- those parameters that influence both the study conditions and the results but are not accounted for and likely not the intended parameters you want to observe. For instance, if you give participants two interfaces with different layouts to use but they involve dozens of unexplained buttons and knobs, the difficulty of the systems will confound the relationship between the instruments' layout and the participants' interaction. The results will be unclear and you will likely learn more about the participants' frustration than their preference for layout.

\textbf{Decide on the Study Setting}: Evaluations of MATs take place in settings ranging from controlled laboratory settings to ad-hoc settings in public. These settings give different levels of control of the study, realism of conducting the activity in-context, and appropriate data collection methods and study structures. Whilst a laboratory study gives the most control, it is the least realistic setting for creative, fun, and engaging interaction. There are trade-offs between different settings, and you should be prepared to justify the decisions you make in planning your study when discussing your research. 

\textbf{Focus the Activity}: Even for creative activities there needs to be some focus to the interaction. Whilst evaluating MATs is not concerned with, say the efficiency or productivity of carrying out specified tasks with an interface, it is nonetheless important to focus peoples’ activity in order to be able to understand their responses to the design’s features. For example, with musical MATs, people need to be provided with a motivation for creating music. This might be in the form of a compositional brief; e.g., in Daisyphone studies, participants were asked to create a jingle for the Olympic Games, and in the Keppi study performers were asked to create a musical performance using the MAT.

\section{Conducting and Reporting Your MAT Study}
\label{ch:methods}

Table \ref{tab:studycomp} outlines the necessary components for conducing proof-of-concept and comparative studies. 
We have structured this section to serve as a template to both guide the design of the research as well as its presentation (e.g., in a research paper). Following this structure and ensuring you have all of the key components in your study planning and presentation of your work will help to make the research rigorous and repeatable, and understandable by other researchers. The template sections you see here are used in the majority of research papers across scientific disciplines. This will help you to structure your presentation in a variety of venues for the topics related to your interdisciplinary work. 


\begin{table}[ht]
\centering
\begin{tabular}{@{}rrcc@{}}
\toprule
\multicolumn{1}{l}{\textbf{}}          & \multicolumn{1}{l}{}           & \multicolumn{1}{l}{\textbf{Proof-of-Concept}} & \multicolumn{1}{l}{\textbf{Comparative}} \\ \midrule
\rowcolor[HTML]{EFEFEF} \multicolumn{2}{r}{\textbf{Background}}                                 & \cmark                                        & \cmark                                   \\ \cmidrule(r){1-2}
\multicolumn{2}{r}{\textbf{Research Questions}}                         & \cmark                                        & \cmark                                   \\ \cmidrule(r){1-2}
\parbox[t]{6mm}{\multirow{9}{*}{\rotatebox[origin=c]{90}{\textbf{Study Methodology}}}} & \cellcolor[HTML]{EFEFEF}\textbf{Aims}                  & \cellcolor[HTML]{EFEFEF}\cmark                                        & \cellcolor[HTML]{EFEFEF}\cmark                                   \\
                                       & \textbf{Hypotheses}            &                                               & \cmark                                   \\
                                       & \cellcolor[HTML]{EFEFEF}\textbf{Independent Variables} &                   \cellcolor[HTML]{EFEFEF}                            & \cellcolor[HTML]{EFEFEF}\cmark                                   \\
                                       & \textbf{Dependent Variables}   &                                               & \cmark                                   \\
                                       & \cellcolor[HTML]{EFEFEF}\textbf{Conditions}            &                   \cellcolor[HTML]{EFEFEF}                            & \cellcolor[HTML]{EFEFEF}\cmark                                   \\
                                       & \textbf{Participants}          & \cmark                                        & \cmark                                   \\
                                       & \cellcolor[HTML]{EFEFEF}\textbf{Tools}                 & \cellcolor[HTML]{EFEFEF}\cmark                                        & \cellcolor[HTML]{EFEFEF}\cmark                                   \\
                                       & \textbf{Procedure}             & \cmark                                        & \cmark                                   \\
                                       & \cellcolor[HTML]{EFEFEF}\textbf{Data Collection}        & \cellcolor[HTML]{EFEFEF}\cmark                                        & \cellcolor[HTML]{EFEFEF}\cmark                                   \\ \cmidrule(r){1-2}
\multicolumn{2}{r}{\textbf{Analysis}}                              & \cmark                                        & \cmark                                   \\ \cmidrule(r){1-2}
\rowcolor[HTML]{EFEFEF}\multicolumn{2}{r}{\textbf{Results}}                         & \cmark                                        & \cmark                                   \\ \cmidrule(r){1-2}
\multicolumn{2}{r}{\textbf{Discussion}}                         & \cmark                                        & \cmark                                   \\ \bottomrule
\end{tabular}
\caption{Components necessary for both types of MAT study.}
\label{tab:studycomp}
\end{table}

\subsection{Section: Background}
The first step should be to conduct a literature review of existing research surrounding your topic area. The related work should create a coherent idea driving the current research. This should be done to help define research questions (e.g., whether there a gap in the existing knowledge or previous publications suggest further exploration you can address in your work) and to provide a state-of-the art for methods being used in similar studies. 
A strong background will help you to justify your choices for study design, methods, and data analysis, thus increasing the robustness of your work. 

You should provide a brief overview of the relevant literature when presenting your research to contextualise what you are doing within an existing body of knowledge and provide rationale for your decisions. With MAT research, the background may consist of relevant work from different disciplines and so you will want to spent time to make sure that you connect literature coming from different fields. This might include defining terminology you will use, with respect to different definitions in other research, or uniting concepts from multi-disciplinary work so you can discuss them together. For instance, in \citet{reed:2021}, the authors use the Background section of the paper to unite concepts surrounding mental imagery and embodiment from cognitive science and design practice within the context of the paper. 

\subsection{Section: Research Questions}
As mentioned in Section \ref{sec:research-qs}, you should define specific research questions you wish to address through your work. With existing literature in mind, these questions will help to focus the research and the design of your MAT and study.
A research question, while targeting a specific area for exploration, is usually something broad such as ``How do people engage with each other when playing a three-person musical instrument?'' (as in Polymetros).
It is common, when writing a research article, to include the research questions and the main contributions your work makes to the existing literature after presenting your Background. This helps to connect the Background to your current work and focus a reader on the key points you will address in the current study.

\subsection{Section: Study Methodology}
Referring to the above table, you should include the following components in your study design and when presenting your research. Methodology is important in order to ensure the study is robust and reproducible. The methodology should include detailed information about the procedure so that another researcher may reproduce it exactly as you did. In your presentation, you may wish to use subsections for each component of the methodology.

\subsubsection{Aims}
You should define and list the aims of the study. There should be a small number of aims – one or two, and definitely less than five.

For a proof-of-concept study, the aim is usually to find out people’s conceptualisation or understanding of a novel piece of interaction, e.g., ``The aim of the study is to explore how people engage with each other when they play a three-person musical instrument.'' (Polymetros).

A comparative study may examine a specific interaction paradigm, perhaps further exploring an observation from a proof-of-concept study. A comparative study example: ``The aim of the study is to test whether a shared music making system with support for shared communication channels supports greater levels of mutual engagement than one without.” (Daisyphone).

\subsubsection{Hypotheses}
If you are doing a comparative study, you will have at least one hypothesis which you test. There should be a small number of hypotheses, usually between one and five hypotheses is fine. For example, in the Daisyphone comparative study there were two hypotheses. The study examined the effect that providing a graphical annotation function as an additional communication channel has on mutual engagement:

\begin{itemize}
    \item[]\textbf{H1}: Mutual engagement would be greater where an additional channel of communication was provided – graphical annotation. (Daisyphone).
\end{itemize}

You might have a more exploratory kind of comparative study and just predict that there would be some difference (but not know what kind of difference it is):

\begin{itemize}
    \item[]\textbf{H2}: Mutual engagement would be different where an additional channel of communication was provided compared to when there is no additional channel of communication.
\end{itemize}

Make sure that, if not already done in the Background, that you define the terminology you will use here; in these hypotheses, you will want to clearly state what is meant by ``mutual engagement,'' and how it might be measured. This should be informed by and connected to your Background section.
For example, in the Daisyphone study ``mutual engagement'' was described as a collaboration in which there is both ``Evidence of engagement with the product of the joint activity, i.e. music in our domain. For example, participants’ reports of feeling engaged with the product, a high quality product, focused contributions, or demonstrations of skills and expertise in creating contributions.'' and ``Evidence of engagement with others in the activity. For example, more reports of feeling engaged with the group, coherent final joint products, colocation of contributions, mutual modification of work, discussions of quality of the joint product, repetition and reinterpretation of others’ contributions'' (as described in \citet{bryankinns:2009}).

\subsubsection{Variables}
Defining variables is an important part of a comparative study plan – variables are things that are changed in your study, or things that are measured in your study (i.e., they are \textit{variable} in the study). There are two main kinds of variables to define: \textit{independent variables} and \textit{dependent variables}. 

You should also be aware of other elements which might act as \textit{confounding variables}. These are things that might change between each participant in your study and which might have an effect on their performance; for instance, the skills and expertise of participants might be a confounding variable and therefore need to be controlled. If some participants are skilled musicians, their experience might have an effect on how they play a 3-person musical instrument. The time of day might be a confounding variable if people are tired in the evening versus the morning, etc. In MAT studies, it is often difficult to control these confounding variables and so they need to be included in the Discussion section of your report and you should elaborate on how they may have impacted the findings and outcomes of your study. As mentioned previously, limiting the complexity of the study will help to keep confounding variables under control.

\paragraph{\textbf{\textit{Dependent Variables}}}
Dependent variables are things that you measure in a comparative study. They \textit{depend} on the interaction and what happens during the study. You should describe these as concretely as possible; for example, in conventional HCI studies, a dependent variable could be the time in seconds it takes to complete a task. For more exploratory MAT studies, your dependent variable might be more subjective; for example, in the Daisyphone study there were 9 dependent variables including: (1) ``Quality measure: participants’ reports of their assessment of the quality of the final product and the collaboration itself,'' (2) ``Contribution to joint production measure: number of notes contributed,'' and (3) ``Proximal interaction measure: closeness of participants’ contributions to each other’s contributions” (as described in \citet{bryankinns:2009}).

For each of these, you will need to define how the dependent variable is measured or how a quality is assessed. The best way to do this is to find good definitions in existing research literature and either use those definitions or modify them to suit your studies.

\paragraph{\textbf{\textit{Independent Variables}}}
Independent variables are things that you change in the study; for example, changing features of the interface or the app which is provided to see their effect on user experience (as measured by your dependent variables). You define the independent variable along with the ``levels'' of variable, or what things you change. For example, to test the effect of providing shared annotation in the Daisyphone study, an independent variable would be ``Annotation'' and the 2 levels would be ``Annotation'' or ``No Annotation'' (Daisyphone): ``In the Annotation condition, participants could `draw' on the Daisyphone, and these graphical annotations were shared with other participants. In the No Annotation condition, no graphical annotation was supported, and so communication could only occur through the music.'' 

\subsubsection{Conditions}
In your comparative study you will test different variants of the interaction -- each of these variants is called a \textit{condition} or treatment in the study.

In the Daisyphone study, two hypotheses were tested: \textbf{H1} ``Mutual engagement would be greater where participants had explicit cues to identity'' and \textbf{H2} ``Mutual engagement would be greater where an additional channel of communication was provided – graphical annotation''.

Two independent variables were used to test these hypotheses: 1) cues to identity (for H1), and 2) communication channel (for H2). There were therefore 4 conditions in this study (see Table \ref{tab:conditions}).

\begin{table}[ht]
\centering
\begin{tabular}{cr>{\centering}p{0.3\textwidth}>{\centering\arraybackslash}p{0.3\textwidth}}
\toprule
\multicolumn{1}{l}{}                  & \multicolumn{1}{l}{} & \multicolumn{2}{c}{\textbf{Independent Variables}}         \\ \cmidrule(l){3-4} 
\multicolumn{1}{l}{}                  &                      & \textbf{Communication Channel} & \textbf{Identity Cues}    \\ \cmidrule(l){3-4} 
                                      & \textbf{1}           &                                &                           \\
                                      & \textbf{2}           & \cellcolor[HTML]{EFEFEF}X      &                           \\
                                      & \textbf{3}           &                                & \cellcolor[HTML]{EFEFEF}X \\
\multirow{-4}{*}{\rotatebox[origin=c]{90}{\textbf{Conditions}}} & \textbf{4}           & \cellcolor[HTML]{EFEFEF}X      & \cellcolor[HTML]{EFEFEF}X \\ \bottomrule
\end{tabular}
\caption{The four conditions examined in the Daisyphone study, showing the combination of the two independent variables. An X indicates where the variable was included in the condition.}
\label{tab:conditions}
\end{table}

\subsubsection{Participants}\label{sec:participants}
Then, decide who your participants will be -- are they general public, or do they need certain skills and experience? If so, what skills and experience? You will also need to decide how you will recruit the participants and if they will be paid any incentives? This information should be described in detail when you document your research. For example, the participants for the Daisyphone study are described in \citet{bryankinns:2009}:

\begin{addmargin}[1em]{2em}
``Final year Computer Science students at the first author's institution were recruited through advertisements to take part in the experiment as part of their course, but not offered any incentives to take part. Thirty-nine of a possible 80 participants took part (28 males, 11 females; aged from 20 to 29 years old, mean age: 22; average computer literacy: expert; average musical ability: intermediate; none were professional or trained musicians; none had used Daisyphone before). Participants’ musical preferences ranged from Hip Hop (most popular) to Latin (least popular).''
\end{addmargin}

If you report individual responses in your results (e.g., interview answers), then you should give a table to provide brief demographic of each participant, as this may have an impact on their answers. Give each person a participant ID (e.g. P1, P2, etc.) which you can then use in the results. For example, see Table \ref{tab:pttable}. 

\begin{table}[ht]
\centering
\begin{tabular}{@{}lllll@{}}
\toprule
\textbf{~~~Participant~~~} & \textbf{~~~Age~~~} & \textbf{~~~Gender~~~} & \textbf{~~~Musical Ability~~~} & \textbf{~~~Musical Preference~~~} \\ \midrule
\rowcolor[HTML]{EFEFEF} P1                   & 25           & M               & Novice                   & Hip hop                     \\
P2                   & 24           & M               & Intermediate             & Hip hop                     \\
\rowcolor[HTML]{EFEFEF} P3                   & 20           & F               & Intermediate             & Hip hop                     \\
P4                   & 22           & M               & Intermediate             & Rock                        \\
\rowcolor[HTML]{EFEFEF} P5                   & 21           & F               & Expert                   & Latin                       \\
P6                   & 22           & M               & Novice                   & Classical                   \\
\rowcolor[HTML]{EFEFEF} P7                   & 22           & M               & Intermediate             & Rock                        \\
P8                   & 21           & F               & Intermediate             & Ambient                     \\ \bottomrule
\end{tabular}
\caption{Example table for presentation of participant demographics and background (adapted from \citep{bryankinns:2009}).}
\label{tab:pttable}
\end{table}

\paragraph{\textbf{\textit{Sample Size}}}
As a rule of thumb, you will need at least 10 participants in a proof-of-concept study. Ideally you would have much more than 10 participants for a proof-of-concept study, but it is worth noting that the most common sample size in papers in the leading HCI conference (ACM CHI) is 12 participants \citep{caine:2016}. For a comparative study, you need at least 10 participants for each condition, e.g. in the Daisyphone example you would need at least 40 participants as there are 4 conditions (assuming that each person only does one condition -- this is called an unrepeated-measure, also referred to as ‘between groups’).

Data suggests that, in order to discover all of the usability problems in testing, 15 participants are needed; through iterative testing, this could be done in as little as 5 participants \citep{Nielsen:93}. You can reduce the number of participants needed in a comparative study by designing the study so that each participant does multiple conditions (\textit{repeated-measures}, also called within-groups), but this makes the study longer (therefore increasing participant boredom and fatigue), increases the chance of learning effects, where participants learn something about the interaction in one condition which then either makes the other conditions easier to use or harder to use due to confusion, and potentially reduces the power of the statistical tests you can do on the data after the study.

Repeated-measures study design is beneficial in that you can ask participants comparative questions between the conditions which can help to understand how participants react differently to different conditions e.g., to compare the experience of condition 1 to their experience of condition 2.

For example, you could run the Daisyphone study with 10 participants and have each person use all four conditions. As another option, you could run the study with 20 participants and 10 of the people use conditions 1 \& 2, and the other 10 participants use conditions 3 \& 4. In this case, you would need to decide why it is not some other combination (e.g., 1 \& 3 + 2 \& 4) and again be prepared to make this justification when presenting the research. When using repeated measures, you also need to make sure to counterbalance the ordering of the conditions so that the ordering does not cause an effect in your results. You can use a balanced Latin square to figure out different orderings for your conditions to ensure that your study is balanced\footnote{Balanced Latin Square Generator: \href{https://cs.uwaterloo.ca/~dmasson/tools/latin_square/}{https://cs.uwaterloo.ca/~dmasson/tools/latin\_square/}}; for instance, if you had a study with four conditions of the Daisyphone study, you could divide your measures as in Table \ref{tab:latin}.

\begin{table}[]
\centering
\begin{tabular}{@{}llllll@{}}
\toprule
                                                            &            & \multicolumn{4}{c}{\textbf{Trial Order}}                                                                            \\ \cmidrule(l){2-6} 
\multicolumn{1}{l}{}                                       & \textbf{1} & \cellcolor[HTML]{EFEFEF}~~~~~~~1~~~~~~~ & \cellcolor[HTML]{EFEFEF}~~~~~~~2~~~~~~~ & \cellcolor[HTML]{EFEFEF}~~~~~~~4~~~~~~~ & \cellcolor[HTML]{EFEFEF}~~~~~~~3~~~~~~~ \\
\multicolumn{1}{l}{}                                       & \textbf{2} & ~~~~~~~2~~~~~~~                         & ~~~~~~~3~~~~~~~                         & ~~~~~~~1~~~~~~~                         & ~~~~~~~4~~~~~~~                         \\
\multicolumn{1}{l}{}                                       & \textbf{3} & \cellcolor[HTML]{EFEFEF}~~~~~~~3~~~~~~~ & \cellcolor[HTML]{EFEFEF}~~~~~~~4~~~~~~~ & \cellcolor[HTML]{EFEFEF}~~~~~~~2~~~~~~~ & \cellcolor[HTML]{EFEFEF}~~~~~~~1~~~~~~~ \\
\multirow{-4.2}{*}{\rotatebox[origin=c]{90}{\textbf{Participant}}} & \textbf{4} & ~~~~~~~4~~~~~~~                         & ~~~~~~~1~~~~~~~                         & ~~~~~~~3~~~~~~~                         & ~~~~~~~2~~~~~~~                         \\ \bottomrule
\end{tabular}
\caption{An example of a balanced Latin square for the four conditions of the Daisyphone study; here, the conditions are in a different order for each participant and each condition precedes another only once, ensuring that potential effects from order are removed from the study.}
\label{tab:latin}
\end{table}

However, it is important to note that you may need more participants depending on what kind of data is being collected and what analyses need to be done. In order to do some statistical testing with accuracy, larger sample sizes might be needed. This should also be considered when recruiting participants and when considering how many variables are examined at one time. 

\subsubsection{Tools: The Media and Arts Technology Itself}

If you need to build a MAT to be used in your study -- for instance, an app, a VR experience, tangible interface, etc., you will need to document the design and be able to explain how it works. This would include detailing your expectation of how people would interact with it, and how it works on a technical level. You will need to connect your design to the Background to be able to define and discuss how it is similar or different to existing tools.
If you are using existing software, describe what the software is 
and how it will be used in this specific research context (making sure to credit the original authors/developers and cite any relevant publications).

\subsubsection{Procedure}
With your materials and tools ready, you should then define the study procedure. Provide a step by step description of what participants in the study will be asked to do, and how long is spent on each step. Include \textit{everything} -- include the step at the beginning when you explain the structure of the study to participants (e.g. 2 minutes), include the part where you ask participants to fill in questionnaire (e.g. 5 minutes), etc. In the pursuit of repeatable studies and reproducible findings, it is important that you are detailed enough for another researcher to conduct the research exactly as you have done.

All studies should start with an introduction where you explain the purpose of the study (but don’t mention what results you are looking for as this may bias their views) and that participants are free to stop at any time they like. You should also collect demographic data such as age, gender, and other relevant information such as musical experience. Specialist skills such as musical experience can be assessed using questionnaires such as the Goldsmiths Musical Sophistication Index (Gold-MSI) \citep{mullensiefen:plosone} -- using existing questionnaires helps to strengthen the rigor of your study. Do not collect anything that could directly identify the participant; for instance, do not collect their name, address, or email address in the demographic information. {If you are studying or working in an institution such as a University you will likely need to ask participants to complete a consent form and ensure that you study has ethical clearance from your institution to proceed.}

\subsubsection{Data Collection}
One critical component of study design is the decision on what kind of data will be needed and collected, either quantitative, qualitative, or some mixture of both. \emph{Quantitative data} is data which is in a numerical form can be assigned a value (eg., it has a quantity --- a numerical value --- and is able to be measured as such). It may be participants' ratings of their enjoyment on a Likert scale from 1 to 5, or it may be more conventional HCI measures such as speed of completing an activity, or number of errors, etc. \emph{Qualitative data} takes the form of words and is more descriptive (eg., it is more about a specific quality or descriptor of whatever is being examined). Many studies incorporate mixed methods, where both quantitative and qualitative approaches are taken. This is especially the case in your MAT research, where you might need to both collect concrete data, for instance on the MAT's computational performance or on the interaction, and also understand the emotional and aesthetic aspects of the design in a qualitative sense. It is important to consider what form your data will take when planning a study to know what kind of analyses you might use to interpret the data, and to ensure that you are able to collect enough data to get an appropriate and well-rounded analysis. As with types of MAT study, one kind of data is not a ``better'' option, but likely one will be better at addressing which perspectives you wish to explore and you should be prepared to justify your choices in your research presentation.

For proof-of-concept studies you will most likely collect data by writing down your observations of people’s interaction with the MAT (and each other), video recording their interaction, and then having questionnaires and interviews at the end of the study to get some idea of how people responded to your MAT. For a comparative study you should describe what data you will collect for each dependent variable. Decide whether the data will be collected whilst the participant is interacting with the MAT, or after they have used it.
Some examples of data collection are outlined below. 

It is important to note any possible ethical considerations with your data collection and how they will be addressed. Check with your affiliated institution's General Data Protection Regulation (GDPR) or other privacy and ethics procedures. For example, if you video record people interacting with your MAT, then does your institution allow you to publish images of people's faces? If so, will you request people’s consent to use their photos in papers? If they don’t give consent, can you still get the data you need for your study? 

\paragraph{\textbf{\textit{Interviews}}}
When conducting interviews with participants, consider what kinds of responses you are seeking. You might use an existing interview structure or a semi-structured interview where you decide prompts based on the topics you wish to explore.
It is best to audio record interviews and then transcribe them later -- people will be much more descriptive and reflective in spoken answers than in written answers. 

Interviews are particularly important for proof-of-concept studies in order to get participants’ feedback on novel experiences. They are also important in comparative studies to help understand why people behaved the way they did and responded to questions the way they did. You should start your interview with open questions, where you try to get participants to explain their understanding of the experience (e.g., ask: ``Please could you describe what you just experienced to me?'', or ``Please tell me how you would describe what you just experienced to a friend?'', followed by more specific question such as ``What did you find most engaging about the system?'', or ``What did you find most challenging about the experience?''). 

Then, you can start to probe for more specific feedback; for example; ``Please could you tell me about your experience of playing music on the three-person music instrument?'', followed by ``Did you find that other people responded to your musical contributions?''. Make sure to follow any questions that could be answered with a yes/ no answer with probing questions, (e.g., ``Can you tell me why that was?'' and/ or ``Can you give me a specific example?''). More specific and targeted questions and answers can help to explain people’s responses to particular features of your MAT.

\paragraph{\textbf{\textit{Questionnaires}}}
It is best to use existing questionnaires, such as the Gold-MSI mentioned earlier \citep{mullensiefen:plosone}, the Creativity Support Index (CSI) \citep{Cherry:2014} or the NASA Task Load Index (TLX) \citep{Hart1988}. These questionnaires are validated through existing research -- this means that they have been shown to be reliable in testing for certain kinds of feedback. When using such questionnaires, it is best not to select a subset of questions -- use the whole questionnaire. You need to be careful to not have too many questions otherwise participants will become bored or frustrated, especially in online studies.

Make sure to choose your questions and questionnaires carefully to address your research questions. You will need to balance your time: not too long for participants to complete without losing focus, and not too short that you don’t get any useful data. As a rule of thumb, you are likely to need at least 10 questionnaire questions to get useful data. Try to restrict the length of questions to one A4 side if possible. In the Polymetros example a 7-question questionnaire was printed on one side of A5 paper, but this was due to very short time for people to complete the questionnaire in a high traffic public venue and resulted in very little useful data.

\paragraph{\textbf{\textit{Video-Cued Recall}}}
You might also ask participants to watch a video recording of their interaction experience and provide a commentary on it. This provides some reflective data on what they did and how they responded to the interaction. To provide results which are more comparable between participants, you might focus on getting participants to identify particular pieces of behaviour you are interested in, (e.g., points at which they learnt a new aspect of the interaction, felt frustrated, introduced new ideas, or they felt most immersed in the experience).

\paragraph{\textbf{\textit{Observations and Retrospective Video Analysis}}}
Observations are usually carried during the interaction to get an overview of the forms of interaction with the MAT. You would then go over video recordings of the interaction to annotate the video with descriptions of the interaction and then code the interaction. For example, you could analyse video in terms of the following coding schemes depending on your study, or develop your own coding scheme:
\begin{itemize}
    \item Participants mirroring, transforming, or complementing each other’s contributions or actions. This is referred to as evidence of Mutual Engagement \citep{bryankinns:2009}.
\item Changes in participants performative interaction - whether they are simply observing, or participating, or performing \citep{sheridan:2008}.
\item Number of new ideas generated e.g., ideas in a brainstorming session.
\item Topics of conversation between participants - what are the conversations predominantly about? The system, the creative act, each other, the organisation of the activity, the weather? If there is a lot of talking between participants, you could use Thematic Analysis (see Section \ref{analysis}) to identify the common themes.
\end{itemize}

\subsection{Section: Data Analysis}\label{analysis}
Data analysis will generally be broken into either quantitative or qualitative analysis. 
Generally, decide and describe what kinds of data analysis you plan to do with the data collected. Often, this will not be perfectly clear before you get the results, but you should provide an indication of the kinds of data analysis you plan to do so that you can collect the right kind of data for the kinds of analysis you want to be able to do. This will be largely dependent on what you want to know from your study. 

\paragraph{\textbf{\textit{Data Measure Types}}}

When collecting data from participants through questionnaires or other feedback mechanisms, you need to specify what kind of measure it is. You will come across the following terms:

\begin{addmargin}[1em]{2em}
Nominal data, or categorical data, uses labels for variables without giving a quantitative value; for instance, having participants indicate gender or using arbitrary categories (e.g., Interface A or Interface B). This data is separated in distinct categories which cannot be ranked.

Ordinal data is similar, but the categories follow a natural order; for instance, asking participants the highest education level they have achieved. There is an order from compulsory schooling to undergraduate to postgraduate studies. This could also include ranking where things labeled as first, second, third preference.
\end{addmargin}

\begin{addmargin}[1em]{2em}
Interval data, or integer data, is measured quantitatively but there is no zero point and it can be negative. The difference between the values on the scale should be measurable and comparable. Age is a common interval data value.

Ratio data is similar but there is a zero-point restricting the range. For instance, asking a participant how many times they use a software during the course of a day. Time duration is also a common ratio data, for example, measuring how long a participant took to complete an activity.
\end{addmargin}

\paragraph{\textbf{\textit{Statistical Testing}}}
For quantitative data, you should decide the statistical tests to perform on the data to see if your observed results match your hypothesis. We use statistical tests to determine whether there are significant relationships between variables or difference between groups. By \textit{significant} here we mean that the differences are not likely to be just due to chance. Ensure that you have enough participants with the variables you are examining, as mentioned in Section \ref{sec:participants}.

You can use \textit{Regression Testing} to check potential cause-and-effect relationships, comparison tests to check for differences or similarities in groups and their behaviour, or correlation analyses to see whether variables are related. When performing statistical analysis, make sure that your data fits into the assumptions of the test being used (if you use a parametric test). Transforming your data can help to achieve normal distribution and variance. A common comparative test used in MAT work is an \textit{Analysis of Variance} (ANOVA); this kind of test examines the difference in two or more groups based on the mean of their data.

\paragraph{\textbf{\textit{Thematic Analysis}}}
With qualitative data, you will need to decide on an evaluation method for the interviews and open-ended questionnaires. Interviews should be transcribed so that you can anonymise and analyse the text data. In MAT research, we commonly use Thematic Analysis on interviews and open-ended questionnaires to identify key themes in people’s responses to MATs \citep{braunclark, braun:2012}. This method is particularly useful for qualitative research as it produces results which are useful for further understanding, rather than qualitative results such as frequency of responses \citep{Braun2020}. For instance, a single participant might make an important comment about their interaction which is different from the others' perspectives and this would be included as an important part of the analysis, rather than as an outlier. You will also need to consider how you will analyse people's behaviour, such as gestures and movement; for instance, coding movement in Laban notation \citep{laban1971mastery, Loke:2005}.

Analyses should be presented referencing the specific method used. This Methods section should only include how the analyses were conducted, saving the results until the subsequent Results section. Here, it is important to describe different measures taken from the data; for instance, how participant involvement was measured, how musical complexity was determined, and so on. 

\paragraph{\textbf{\textit{Triangulation}}}
Additionally, the Methods section is an ideal place to discuss why different analysis methods were were chosen; for instance, justifying the choice of a statistical test. If you want to know about more users' qualitative perceptions of their interaction, it is probably not necessary to conduct statistics testing (and you may not have data which you could test in this way). Therefore, analysing participant responses is more appropriate. If you want to know about more quantitative differences between one interface or another, you could use statistical tests; for instance, seeing if there are more annotations made on one kind of interface than the other (as with Daisyphone). 

In most studies, you will want to have a bit of both kinds of data -- statistical results are used to support the qualitative observations made, and qualitative results are used to contextualise the numerical data gathered. This practice is referred to as \textit{triangulation} --- triangulation is an analysis practice which uses multiple analysis techniques and data streams to get a clear, robust, and well-rounded picture of the study\footnote{Note that, although it is called \textit{triangulation}, this does not mean that specifically three methods must be used}. In the Daisyphone example, quantitative information was collected about the participant's annotation behaviour (number of annotations, locations marked, timing of annotation, etc.) through the program itself. Interviews were also used to gather open-ended response data about the participants' experiences with Daisyphone. Together, the triangulation of these results tell us not only what was going on, but also why. For instance, qualitative analyses of survey responses after working with Polymetros, which suggest controllability and appropriate interest by broader participant groups, was highly associated with participants' responses of feelings of being in control and enjoying the collaborative creative process. \cite[pp.~238-239]{bengler:2013}.

\subsection{Section: Results}
Results should be presented based on the types of analyses done. If you performed multiple analyses, for instance a statistical analysis of study outcomes and then a Thematic Analysis of participant feedback, the results section should be divided into separate parts. This can also help to explain the triangulation of the analysis, making sure to explain each approach and analysis separately before connecting them in the subsequent Discussion section.

\subsubsection{Reporting Statistical Tests}
Depending on which format your publication is in, the presentation of statistical results will be slightly different, but they generally follow the same formats (make sure you check the formatting for the style of your presentation). For instance, with an ANOVA we would report something like:
\begin{itemize}
    \item[] \textit{F}(between group DoF\footnote{Degrees of freedom, the maximum number of independent values than can logically occur in your dataset}, within group DoF) = the F statistic, p = p-value
    \item[] Example: $F(2,26) = 8.76, p = .012$.
\end{itemize}

\paragraph{\textbf{\textit{The p-value}}}
The p-value is the probability that the result of the statistics test was due to chance. In HCI, we usually consider something to be statistically significant if the $p$ is less than .05 ($p < .05$)\footnote{NB: In APA formatting, you should not include a leading 0 when reporting p-values because they exist only between 0 and 1.}. You will also see other p-values of $< .01$ and $< .001$ – these are not typically used in HCI research (see below for more information on the confidence interval).

It is worthwhile to mention that there is some debate about p-values being viewed as the end-all-be-all in results reporting. When you examine the results of your statistics tests, it is important to use critical thinking about your analyses and interpret the p-value in an appropriate way. As mentioned before, your research will be much stronger if you can connect the results of your statistics testing to other data you collected -- in a way, the statistics results are meant to support the validity of the observations you made in your study (e.g., in the Polymetros example of triangulation \citep{bengler:2013}).

\subsubsection{Questionnaire Results}
Generally, when reporting questionnaire results, you want to report either a data count or a range-mean-standard deviation set. What you report depends on the kind of data. For nominal and ordinal data, you will generally want to report a count; for example, ``All participants indicated the interface was Enjoyable (18 participants) or Very Enjoyable (30 participants) to use.'' For interval and ratio data, you will want to include the range of the response, the mean, and the standard deviation; for instance, ``Participants reported that they used the new interface for a longer duration of time, ranging between 45-65 minutes (M = 50, SD = 2.5).''

\begin{figure}[ht]
    \centering
    \includegraphics[width=0.7\textwidth]{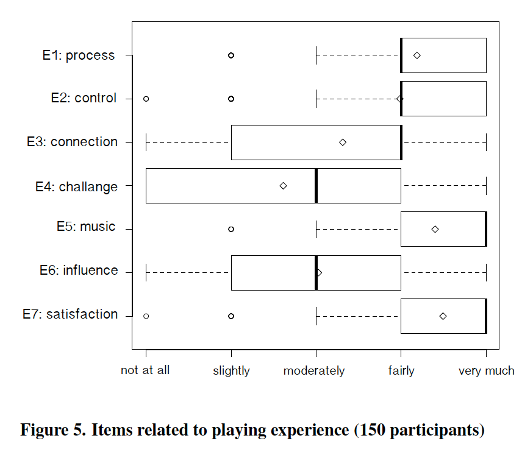}
    \caption{Presentating results using a box-and-whisker plot (reproduced from \citet{bengler:2013}).}
    \label{fig:box}
\end{figure}

You can visually report the results of Likert scale questions using a box-and-whisker plot. These are good at showing many aspects of your data – they show the minimum, the maximum, the median, the upper quartile (top 75\%) and lower quartile (lower 25\%) of your results. For example, the box-and-whisker plot in Figure \ref{fig:box}, referenced from the Polymetros study, shows responses given for elements of playing experience \citep{bengler:2013}. 

\paragraph{\textbf{\textit{Participant Quotes}}}
It is often helpful to provide quotes from participants to illustrate your results, especially when reporting the results of questionnaires and interviews. This provides evidence that you have objectively collected opinions about your MAT. In addition, referencing specific data points and codes when describing a theme from a thematic analysis can help to better contextualise the material and the meaning of the theme. When referencing specific points, make sure to include the participant's ID number.

Quotes should be short and concise to support specific points, only choosing the relevant portion of the feedback, e.g., if a participant said or wrote in their response: ``Well, I enjoyed the experience a lot and found it quite engaging. I was thinking about it the other day whilst walking to work and thought that it was quite nice. That is just my opinion of course, but in general that is my feeling about it. Yes.'' (P2), you should not quote the whole paragraph, but instead quote the key points, e.g., ``I enjoyed the experience a lot and found it quite engaging'' (P2). The other points are just repeating this point. If you are really short of space, you can reduce it further, e.g., ``enjoyed the experience a lot... quite engaging'' (P2).

Make sure to include a range of participants’ responses -- don’t just report one or two participants. If you have 5 participants, make sure to provide quotes from each of them in your results. If you have more than 5 participants then try to report a balance of responses -- positive as well as negative. Remember that negative responses are still useful in terms of research, and give you something interesting to discuss in your discussion section e.g., to discuss why you think that they gave negative responses etc.

\paragraph{\textbf{\textit{Codes from Thematic Analyses}}}

When reporting a qualitative method such a Thematic Analysis (TA), 
it is useful to include the level of detail which was transcribed and analysed (e.g., did you include non-verbal utterances such as laughter, facial expressions, etc.).

Report how many codes you developed in the analysis and provide a general overview of the themes (you do not need to list the codes, but this may be helpful in getting a clearer view of the data). List the themes that you identified in the TA. You can also list the number of codes within each theme to give an idea of how many times the theme was found in participant responses:

\begin{addmargin}[1em]{2em}
E.g., “The 949 coded segments were clustered into the codes: effort; entanglement; characteristics of the compositions; reflections on the instrument; gestures and techniques; performing perception; performer’s body; movement; learning the instrument over time; and ‘edge-like interactions’” \cite[p.~5]{mice:supersize}.
\end{addmargin}

Then provide a description of each of the themes. The description is usually at least a couple of sentences, and if you have space it could be several paragraphs. In the description make sure to illustrate your description with quotes from participants and refer to the participant ID. Remember that TA is about providing a cohesive picture of the data from the codes, so the themes should be linked together and discussed thoroughly. For instance, the \textit{performer's body} theme from Mice \& McPherson is further elaborated as:

\begin{addmargin}[1em]{2em}
E.g., “During the course of the study, we noticed examples of participants feeling differently about their bodies while performing the instrument. Some comments were overwhelmingly positive, for example P5 said the instrument makes her body feel powerful, while other comments implied that participants would like to change their bodies to be more suitable for performing such an oversized instrument. P5 said performing the instrument “makes me want more arms”, and P8 commented “I need bigger arms”. P9 said “I wish I had 3 hands”.” \cite[p.~12]{mice:supersize}.
\end{addmargin}

Remember that the number of codes within a theme does not dictate relevance; the themes are meant to create a clear picture of the data and capture important points and similarities. If the theme contains only one or two codes but provides a critical observation from the research, it is still a valid theme \citep{Braun2020}. With subjective research, participant responses when working with MATs will often demonstrate a wide diversity of interaction perspectives and techniques, and it is important to give attention to this variability; for instance, participant responses when working with the Keppi \cite[pp.~49-50]{bin:2018} or Chaos Bells \cite[pp.~9-12]{mice:supersize} are discussed in detail to represent the varying viewpoints when working with the instruments.

\subsubsection{Other Observations}
It may be beneficial to provide other observations made during the study which do not fall into a specific category of analysis. For example, providing vignettes of observations that support or illustrate findings from your other data. Vignettes are short descriptions of some action and interaction and would often be accompanied by an image from your video recording if you have it.

\begin{figure}[ht]
    \centering
    \includegraphics[width=\textwidth]{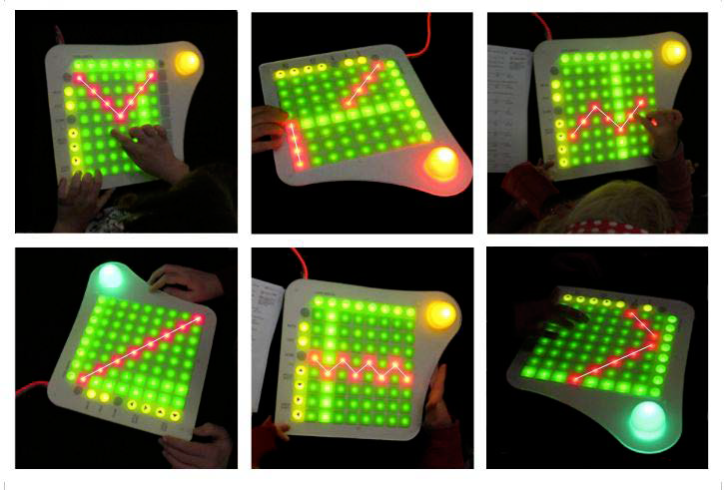}
    \caption{A vignette of typical contributions with Polymetros (reproduced from \citet{bengler:2013}).}
    \label{fig:contributions}
\end{figure}

\begin{addmargin}[1em]{2em}
E.g., from \citep{bengler:2013}:
``A prevalent input strategy was the creation of musical patterns characterised by simple geometric properties. The most common phrases consisted of horizontal and upward or downward diagonal lines whereas in most cases all available notes were used (Figure \ref{fig:contributions}). Resulting in ‘closed musical figures,’ this approach was applied by many players providing a clear audio-visual correlation between the representation on the interface and the musical result.''
\end{addmargin}

\subsection{Section: Discussion}
In the Discussion, you should reflect on your results -- what results were unexpected, why do you think that might be? What explanations can you think of for the results you found? Connect your results back to the Background and the aims of your study -- did your results match with the results of other researchers you mentioned in the literature review? If not, why not? What do your results tell you about the things you wanted to find out in the aims of your study?

It is quite usual to find things that you did not expect, or which are surprising and counter intuitive. This is part of why we conduct these exploratory MAT studies. The Discussion can explore why the results were unexpected and suggest future work to further explore the findings in future work (hopefully using your well-reported and rigorous methods). The most important thing is to clearly and objectively report the results, and then reflect on your results, connecting them to your literature review, your study aims, and your intuition about what happened.

\subsubsection{Limitations}
The Discussion is also the place to discuss limitations to your research. In terms of \textit{internal} and \textit{external validity}. Ideally, your research will have both; in the case it does not, you will need to highlight where the reported results are limited.

Internal validity is the extent to which the study accurately and confidently depicts the relationship between the variables being examined. Some factors which might influence the internal validity are uncontrolled or confounding variables, small sample size, repeated testing and learning effects and potential experimenter bias.

External validity refers to the the extent that the findings of the study can be applied in other settings -- how generalisable the findings are to the world outside of the study. Some factors which might reduce the external validity are selection bias (e.g., the participants who take part in a MAT study might be interested in or already using technology in artistic settings, so might react differently than others who are not as likely to participate), situational factors such as time of day and location, and limited examined factors, such as only looking at music within the Western canon.

In either case, you must identify these factors in validity, not only to lead future work but also to demonstrate your awareness and acknowledgement of potential limitations in the work. For instance, in \citet[pp.~8-9]{reed:2021}, the autoethnographic approach means that, while the results are useful for design and interaction research, the specific interactions discussed in the paper may not be generalisable to other singers with their individual perspectives:

\begin{addmargin}[1em]{2em}
``It is of course critical to again state that, while the interaction observed provides a detailed account of a prolonged interaction with biofeedback through sEMG, this interaction is highly specific to the user... As mentioned previously, further studies to conduct similar trials and autobiographical use cases with the system will be necessary to validate the universality or differences in the experiences.''
\end{addmargin}

\subsubsection{Future Work}
After providing the results and having discussed some limitations, it is worthwhile to suggest future research studies which will further explore the results you have achieved or address the limitations. You might also include some suggestions for research based on interesting or exciting findings, to expand or apply what was learned through the study to additional research questions.

\subsection{Other Presentation Components}
When presenting your research, you will likely need to include an Abstract, Introduction, and Conclusion to bookend the components outlined here. These portions of a paper are sometimes written last, after all of the other information is in place, to better summarise everything in the presentation together.

After the main body of the paper, you may also wish to present your materials alongside your research and results in an appendix. This will help with reproducibility and allow others to follow your methods in their own work. In your appendix you could therefore include all questionnaires, exactly as given to participants and lists of interview questions. 

\section{Conclusion}

Research in Media and Arts Technology (MAT) must strike a careful balance between arts and science practices and norms. This can be a difficult and yet rewarding balance to achieve.
However, the two are not mutually exclusive and the interaction between HCI, User Experience, and Interactive Arts and Media have led to the advancement of both fields. While scientific approaches make the research grounded and extendable to other fields, arts approaches offer new inspiration, contextualisation, and opportunities to explore many facets of the human condition. Through this guide, we have introduced you to scientific practices to make the design of MATs, their study and evaluation, and their presentation and dissemination to others both rigorous and repeatable. Through this guided approach, we hope to increase the accessibility and validity of MAT research in wider artistic and scientific communities and to further explore the human condition of being in the world.

\section*{Acknowledgments}
Work on this chapter was supported by the EPSRC and AHRC Centre for Doctoral Training in Media and Arts Technology (EP/L01632X/1).
We would like to thank all the students and staff at the Media and Arts Technology Centre and the Centre for Digital Music at Queen Mary University of London who have contributed to ongoing discourse about the nature of Media and Arts Technology which inspired many of the thoughts in this chapter.

\bibliographystyle{apalike}
\bibliography{refs}

\end{document}